\documentclass[10pt,twocolumn,floatfix]{revtex4}

\usepackage{epsfig}
\usepackage{amsmath}
\usepackage{amssymb}
\usepackage{mathrsfs}
\usepackage{float}
\usepackage{braket}
\usepackage{hyperref}
\usepackage{dcolumn}
\usepackage{bm}
\usepackage{bbm}
\usepackage{graphicx}
\usepackage{wrapfig}
\usepackage{subfigure}

\newcommand{\p}{\partial}

\renewcommand{\Re}{\mathrm{\mathop{Re}}}
\renewcommand{\Im}{\mathrm{\mathop{Im}}}
\newcommand{\sket}[1]{\ket{#1\rangle}}
\newcommand{\Id}{\mathbbm{1}}

\newcommand{\Op}{\Omega_{\rm{p}}}
\newcommand{\Os}{\Omega_{\rm{s}}}
\newcommand{\Dp}{\Delta_{\rm{p}}}
\newcommand{\Ds}{\Delta_{\rm{s}}}
\newcommand{\gp}{\gamma_{\rm{p}}}
\newcommand{\gs}{\gamma_{\rm{s}}}
\newcommand{\Ep}{E_{\rm{p}}}
\newcommand{\Es}{E_{\rm{s}}}
\newcommand{\nup}{\nu_{\rm{p}}}
\newcommand{\nus}{\nu_{\rm{s}}}
\newcommand{\Lp}{L_{\rm{p}}}
\newcommand{\Ls}{L_{\rm{s}}}
\newcommand{\cp}{\chi_{\rm{p}}}
\renewcommand{\i}{\mathrm{i}}
\newcommand{\Lio}{\mathcal{L}}
\newcommand{\figref}[1]{figure~\ref{#1}}

\newcommand{\secref}[1]{Section~\ref{#1}} 

\begin{document}

\title{Exact Analytical Solution of the Driven Qutrit in an Open Quantum System:\\ V and $\Lambda$ Configurations}

\author{Zachary C. Coleman$^{1,2}$ and Lincoln D. Carr$^2$}

\affiliation{$^1$Natural Science Department, Pepperdine University, Malibu, California 90263, USA}

\affiliation{$^2$Quantum Engineering Program and Department of Physics, Colorado School of Mines, Golden, Colorado 80401, USA}

\begin{abstract}


We obtain the exact analytical solution for the continuously driven qutrit in the V and $\Lambda$ configurations governed by the Lindblad master equation.  We calculate the linear susceptibility in each system, determining regimes of transient gain without inversion, and identify exact parameter values for superluminal, vanishing, and negative group velocity for the probe field.
\end{abstract}

\maketitle

\section{Introduction}
The driven qubit, or 2-level system, subject to decoherence via contact with a Markovian bath is a well-known, exactly analytically solvable problem in quantum computing~\cite{sziklas1969decoherence, schenzle1976twolevel}, including e.g.~T1 and T2 decoherence times~\cite{zhou2008relaxation}.  However, despite the known advantages for quantum computing with qutrits~\cite{randall2015qutrit,low2020qutrit,wu2020qutrit}, the 3-level system in a similar configuration has only been solved approximately.  These approximate investigations of the qutrit have nonetheless revealed many intriguing and unconventional phenomena brought about by quantum coherence and interference not witnessed in the qubit~\cite{scullyzubairy}.  For example, in coherent population trapping (CPT), the system becomes trapped in a ``dark state'' superposition of states which cannot be excited by incident light~\cite{arimondo1996CPT,vanier1998CPT}.  This paradigm is one of several that give rise to lasing without inversion (LWI)~\cite{harris1989LWI,bergou1991LWI,kocharovskyaya1992LWI,mompart2000LWI}, in which light amplification can be achieved despite the ground state population exceeding that of the excited states.  Underlying these effects is the phenomenon of electromagnetically induced transparency (EIT)~\cite{harris1990EIT,fleischhauer2000EIT,fleischhauer2005EIT}, in which a light pulse can be transmitted through an otherwise opaque medium that is effectively rendered transparent.  Experimental demonstrations of these effects~\cite{wynands1999CPT,xiao1995experiment,zibrov1995experiment,kitching1999experiment} have testified to the utility of quantum coherence generated in 3- and higher-level systems.  Although these particular results have been successfully predicted and demonstrated without the exact solution for the dynamics of the driven qutrit in contact with a reservoir, the advantages of qutrits in quantum computing reveal the desirability of this solution.  Here, we develop a method for exactly solving the dynamics of the doubly-driven qutrit in contact with a reservoir as modeled by the Lindblad master equation and report solutions for the V and $\Lambda$ configurations in particular.

\begin{figure}[t]
    \centering
    \subfigure[]{
    \includegraphics[width=.45\columnwidth]{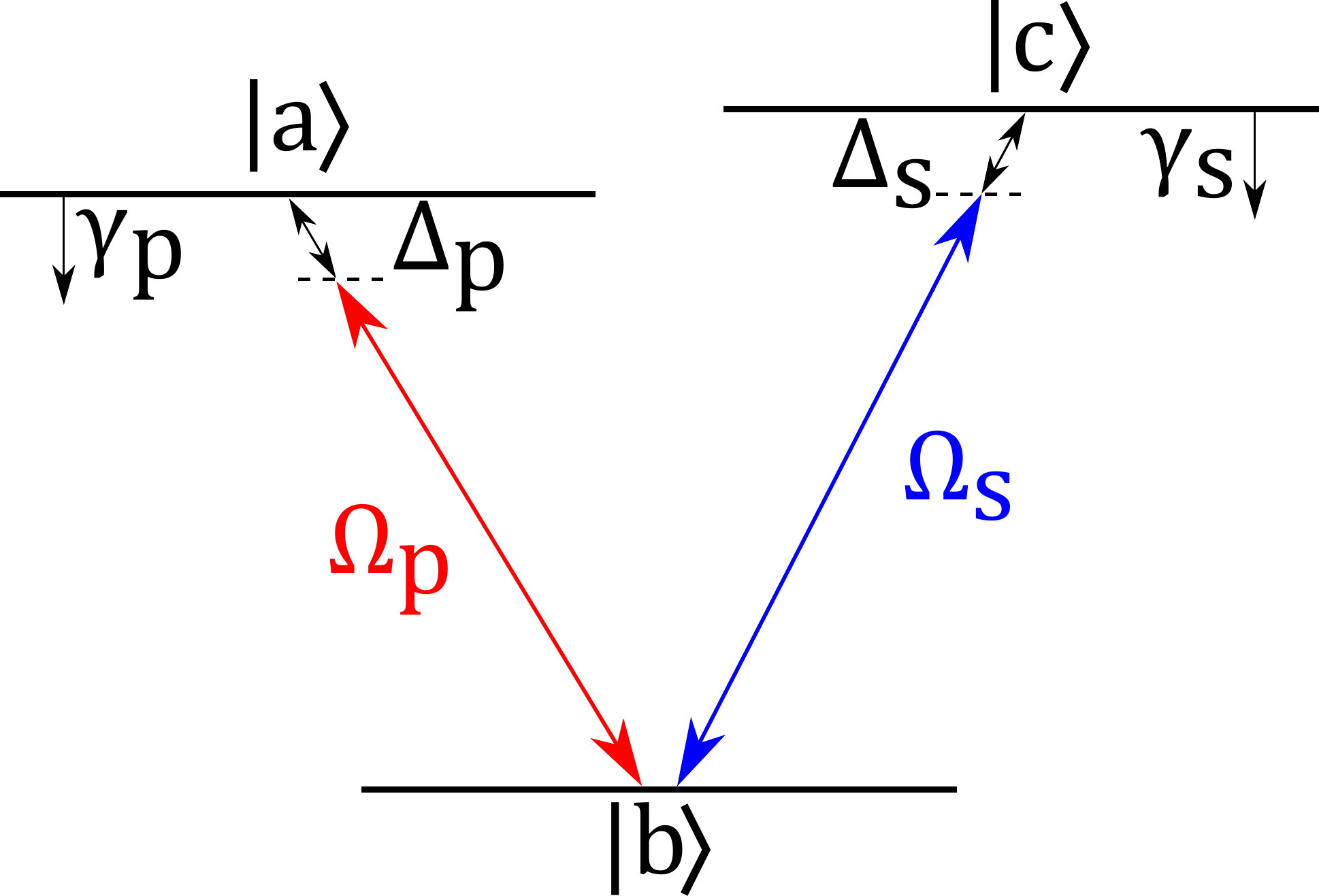}
    }
    \subfigure[]{
    \includegraphics[width=.45\columnwidth]{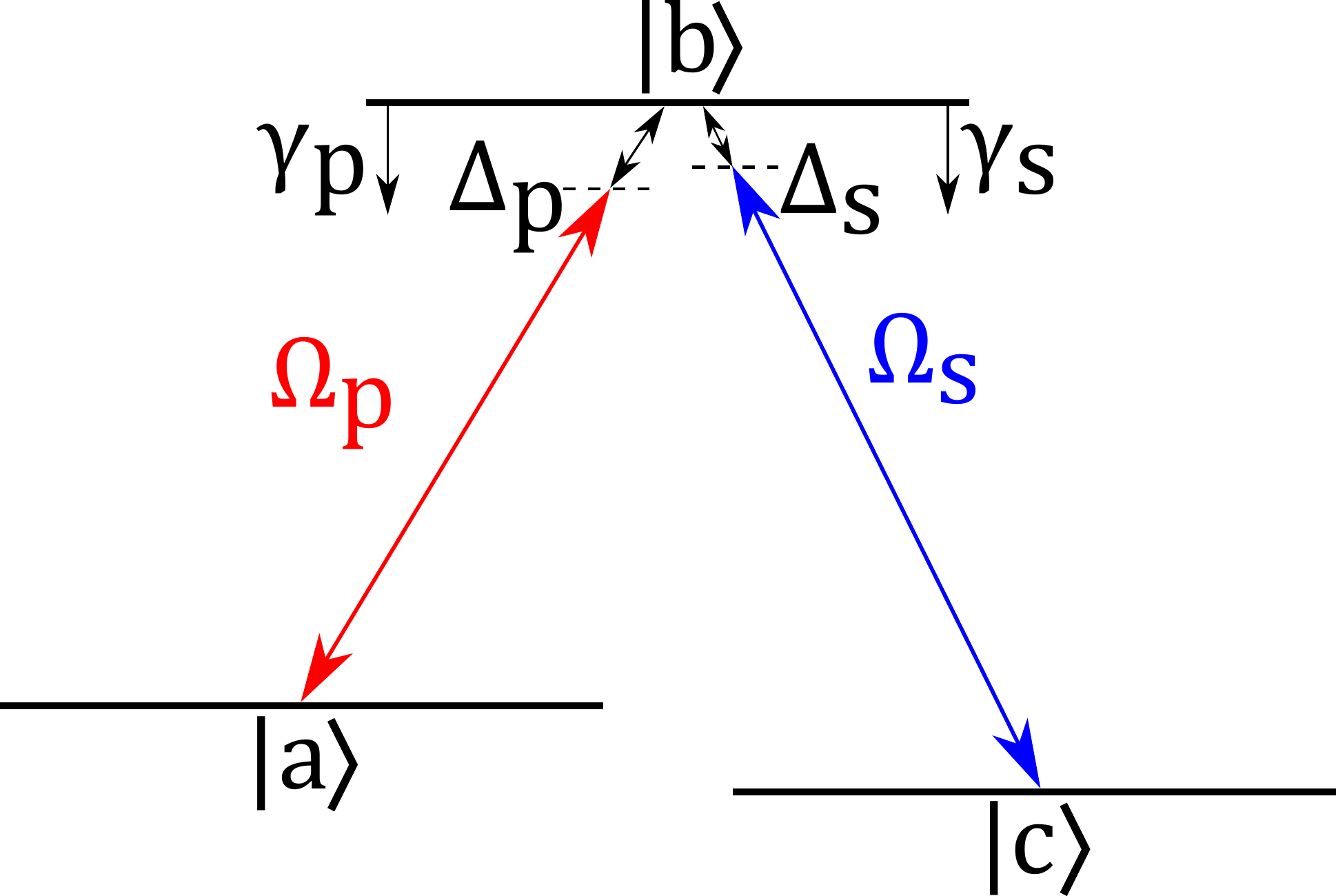}
    }
    \caption{The (a) V and (b) $\Lambda$ configurations for the 3-level system.  In the V-system, the lower state $\ket b$ is coupled to upper states $\ket a$ and $\ket c$, which decay to the ground state with rates $\gp$ and $\gs$ respectively.  In the $\Lambda$-system, the lower states $\ket a$ and $\ket c$ are coupled to the upper state $\ket b$, which decays to each lower state.  In each configuration, the $\ket a\leftrightarrow\ket b$ transition is driven by a probe field with Rabi frequency $\Op$, and the $\ket b\leftrightarrow\ket c$ transition is driven by a control field with Rabi frequency $\Os$.}
    \label{schemes}
\end{figure}

The V- and $\Lambda$-systems  we analyze are schematically shown in \figref{schemes}.  We take the open dynamics of these systems to be governed by the Lindblad master equation, making the Markov and secular approximations relevant  e.g.~for large reservoirs.  We then transform the Lindblad master equation into superoperator form~\cite{schaller2014oqs}.  This allows us to solve for the time dependence of each element of the density matrix.  Using this density matrix solution, we explore regimes of transient and steady-state absorption and light propagation through a calculation of the linear susceptibility.  

Most schemes for achieving LWI and EIT rely on more complex multilevel systems than we consider here~\cite{zibrov1995experiment,kitching1999experiment}.  Theoretical consideration of these systems, however, necessitates many approximations or a numerical approach.  Even in strictly 3-level systems, the usual approach for deriving the EIT phenomenon involves finding steady-state conditions for equations of motion obtained within approximations of the zeroth order in atomic populations or of the first order in the probe field~\cite{scullyzubairy,fleischhauer2005EIT}.  These approximations only hold on small time scales and therefore the solutions they yield differ vastly from our results.  More generally, one may apply techniques of quantum regression~\cite{schaller2014oqs} or perturbation theory~\cite{lazoudis2011EIT}, both of which require strict approximation unlike the method we develop here.

{Related problems to quantum decoherence, such as photon dissipation, have been studied using exact analytical methods~\cite{chen2017analytical}, entanglement-preserving methods~\cite{chen2018entanglement}, and computational methods~\cite{chen2017computational}.  Exact results, which afford arbitrary precision, can have major advantages over numerical approaches, especially in high precision applications such as quantum information.  In particular, the analytical results we develop here offer insight into the physics of qutrits interacting with environments, enabling, e.g.,~expansions in key parameters to arbitrary order.  Although our exact results are obtained within the Lindblad approximation, this also enables exact quantification of a system's deviation from Lindblad conditions, leading to an improved understanding of when and to what extent the Lindblad approximation is valid.}

This paper is organized as follows: In \secref{sec:analytics}, we develop a method for solving the general qutrit; in \secref{sec:V} we present the solution of the V-system along with relevant physical implications; in \secref{sec:L}, we do the same for the $\Lambda$-system.  Finally, in \secref{sec:conc}, we make some concluding remarks on the potential applications of our findings.

\section{Analytical Treatment}\label{sec:analytics}

We develop here an analytical framework to exactly solve the dynamics of the density matrix for the doubly-driven qutrit and apply this method to the V- and $\Lambda$-systems.  
For each of these systems, we allow the $\ket a\leftrightarrow\ket b$ and $\ket c\leftrightarrow\ket b$ transitions to be driven by a probe field with Rabi frequency $\Op=|\mu_{ab}|\Ep/\hbar$ and control field with Rabi frequency $\Os=|\mu_{cb}|\Es/\hbar$ respectively.  Here, $\mu_{ab}$ and $\mu_{cb}$ are the dipole matrix elements associated with each transition, and $\Ep$ and $\Es$ are the amplitudes of the driving fields.
The fields have angular frequencies of $\nup$ and $\nus$ which are detuned from the resonant frequencies of the transitions $\omega_{ab}$ and $\omega_{cb}$ by amounts $\Dp=\omega_{ab}-\nup$ and $\Ds=\omega_{cb}-\nus$.
In the V and $\Lambda$ configurations, the $\ket a\leftrightarrow\ket c$ transition is dipole forbidden.
Additionally, to represent the open dynamics of the system, we introduce decay rates $\gp$ and $\gs$ along each transition due to environmental coupling, with directions appropriate to the energy configurations.  See \figref{schemes} for a level scheme of each system.  Though a weak probe limit is generally taken (i.e.~$\Op\ll\Os$)~\cite{lazoudis2011EIT}, we make no such restriction here.

We obtain the time-independent Hamiltonian for the V- and $\Lambda$-systems within the RWA and through a transformation into a co-rotating frame:
\begin{equation}\label{ham}
	H=\frac\hbar2\begin{bmatrix}\pm2\Dp & -\Op & 0\\ -\Op & 0 & -\Os\\ 0 & -\Os & \pm 2\Ds\end{bmatrix}.
\end{equation}
For the V-system, diagonal elements are positive, while for the $\Lambda$-system, they are negative.

Making the Markov and secular approximations, we take the dynamics of the system to be governed by the Lindblad master equation~\cite{schaller2014oqs}:
\begin{multline}\label{lindblad}
	\frac{\p\rho(t)}{\p t}=-\frac{\i}{\hbar}[H,\rho(t)] \\
		 -\sum_{n=p,s}\frac{\gamma_n}{2}\left(\{L_n^\dagger L_n,\rho(t)\}-2L_n\rho(t) L_n^\dagger\right),
\end{multline}
where $\Lp$ and $\Ls$ are Lindblad operators corresponding to decay transitions.  In the V-system, we use the operators $\Lp=\ket{b}\bra{a}$ and $\Ls=\ket{b}\bra{c}$, while in the $\Lambda$-system, we use $\Lp=\ket{a}\bra{b}$ and $\Ls=\ket{c}\bra{b}$.
In order to convert~\eqref{lindblad} into a tractable form, we perform a transformation into superoperator space~\cite{manzano2020lindblad}, in which our $3\times 3$ density matrix transforms into a $9\times 1$ superket by $\rho(t)\to\sket{\rho(t)}$.
Defining the Liouville superoperator $\Lio$ by the equation $\Lio=-\i/\hbar\cdot(H\otimes \Id-\Id\otimes H^\top)-\sum
\frac{\gamma_n}{2}(L_n^\dagger L_n\otimes\Id + \Id\otimes (L_n^\dagger L_n)^\top - 2L_n\otimes (L_n^\dagger)^\top)$, where $\Id$ is the $3\times 3$ identity matrix,
the general solution to~\eqref{lindblad} in superoperator space is
\begin{equation}\label{soln}
	\sket{\rho(t)}=\exp\left(\Lio t\right)\sket{\rho(0)},
\end{equation}
where $t=0$ is chosen to be the initial time.  We can bring this solution into a closed form by performing the Laplace transform~\cite{schaller2014oqs} on $\sket{\rho(t)}$, which yields the density matrix $\sket{\tilde\rho(s)}$ in the complex frequency domain as
\begin{align}
	    \sket{\tilde\rho(s)}&=\int_{0}^\infty \sket{\rho(t)}e^{-st}dt\notag\\
		    &=\left(s\cdot\Id-\Lio\right)^{-1}\sket{\rho(0)}.\label{lap}
\end{align}
Here, $\Id$ is the $9\times 9$ identity matrix.
We perform the matrix inversion in~\eqref{lap} symbolically, then take the inverse Laplace transform to return the density matrix to the time domain.  We perform the transform on each nonzero term by finding the residue about each of its poles and employing Cauchy's residue theorem.  This method allows for analytic computation of all elements of the density matrix.

By proper modification to the Hamiltonian in~\eqref{ham} and the Lindblad operators in~\eqref{lindblad}, one could apply this method to the qutrit with any decay structure under any time-independent Hamiltonian.  However, the fully general case involves solving for the roots of a ninth-degree polynomial, and the solution may elide a closed form.  Nonetheless, by restricting ourselves to only the V and $\Lambda$ configurations and making the simplifications $\Dp=\Ds=\Delta$ and $\gp=\gs=\gamma$, with the system initially occupy the $\ket b$ state, we reduce the problem to solving third and fourth degree polynomials for the V- and $\Lambda$-systems respectively.  Thus, in all following discussion and plots, $\Delta$ represents the simultaneous detuning of both fields, and $\gamma$ represents the decay rate of both transitions.  These simplifications allow us to present the solutions for these systems in a closed form.  Other simplifications may similarly allow for closed-form results for the 3-level ladder system or for higher level systems, such as the N-type four-level system, but we leave these question for future consideration.

\section{\texorpdfstring{V-System}{Vee-System}}\label{sec:V}

\subsection{Analytical Solution}
For the V-system initially in the $\ket b$ state, we find the following results for the time-dependence of the population elements of the density matrix:
\begin{subequations}
\begin{align}
    \rho_{aa}(t)&=\frac{\Op^2}{\Omega^2}+\sum_{k=1}^3\left(\frac{A_k\exp(\lambda_k t)}{\prod_{l\neq k}^3(\lambda_k-\lambda_l)}\right)\label{Vaa}\\
    \rho_{cc}(t)&=\frac{\Os^2}{\Op^2}\rho_{aa}(t)\\
    \rho_{bb}(t)&=1-\rho_{aa}(t)-\rho_{cc}(t),
\end{align} 
with the definitions
\begin{align}
    \Omega&=\sqrt{4\Delta^2+\gamma^2+2(\Op^2+\Os^2)}\label{VOmega}\\
    A_k&=\Op^2(\gamma+2\lambda_k)/4\lambda_k,
\end{align}
\end{subequations}
and the $\lambda_k$ as the three roots to the polynomial $(\gamma+\lambda)((\gamma+2\lambda)^2+4\Delta^2)+2(\gamma+2\lambda)(\Op^2+\Os^2)$.  These roots can be solved exactly, but their expressions are unwieldy, so we do not reproduce them here.  We additionally find the coherence elements
\begin{subequations}
\begin{align}
    \rho_{ab}(t)&=\frac{\Op}{\Omega^2}(2\Delta+\i\gamma)-\sum_{k=1}^3\left(\frac{B_k\exp(\lambda_k t)}{\prod_{l\neq k}^3(\lambda_k-\lambda_l)}\right)\\
    \rho_{bc}(t)&=\frac{\Os}{\Op}\rho_{ab}^*(t)\\
    \rho_{ac}(t)&=\frac{1}{2\Op\Os}(\Os^2\rho_{aa}(t)+\Op^2\rho_{cc}(t)),\label{Vac}
\end{align}
with\begin{equation}\label{bkV}
    B_k=\i\Op(\gamma+\lambda_k)(\gamma+2\lambda_k-2\i\Delta)/4\lambda_k.
\end{equation}
\end{subequations}
Note that each term of the density matrix consists of a single constant term and several time-dependent terms.  In the steady state, the time-dependent terms vanish (as $\Re(\lambda_k)<0$ for each $k$), and we are left with only the constant term.  The density matrix in the steady-state represents a mixed state with $\mathrm{\mathop{Tr}}(\rho^2)<1$.

We now compare our exact results with previous approximate results where available.  We note a conservation law between the excited states and the coherence between them, expressed as
\begin{equation}\label{conservation}
	\rho_{aa}(t)+\rho_{cc}(t)-\frac{\Op^2+\Os^2}{\Op\Os}\rho_{ac}(t)=0.
\end{equation}
In the case of matching driving parameters, i.e., $\Op=\Os$, \eqref{conservation} reduces to results in~\cite{kozlov2006coherence}, where a factor of $2$ appears in the third term, though Kozlov et al.~consider incoherent pump rates in contrast to our coherent driving fields.

\begin{figure}
    \centering
    \includegraphics[width=\columnwidth]{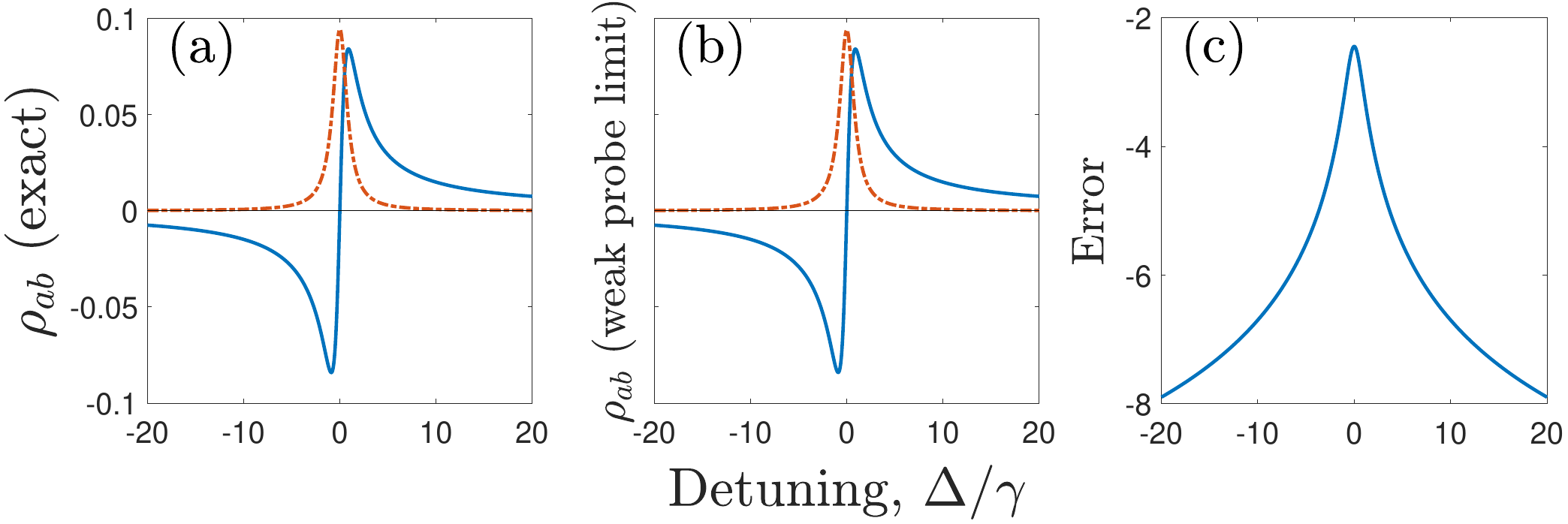}
    \caption{(a) Exact solution of the real (solid) and imaginary (dashed) parts of the probe coherence terms $\rho_{ab}$ in the steady state, along with (b) the same term expanded to first order in $\Op$ (i.e. the weak probe limit), and (c) the error between the two, calculated as $\log|(\rho_{ab,exact}-\rho_{ab,appx.})/\rho_{ab,exact}|$.  The weak probe limit performs strongly even when $\Op$ is less than a full order below $\Os$.  Here and in all following plots, $\Delta$ represents the simultaneous detuning of both the probe field and the control field. Parameters: $\Op=3\times10^5$, $\Os=10^6$, $\gamma=10^6$.}
    \label{weakprobe}
\end{figure}

In the steady state, under conditions that $\Delta=0$ and $\gamma\ll\Op,\Os$, we find as a zeroth order approximation that for any driving intensity, half of the population will become locked in an excited state, with the ratio $\rho_{aa}/\rho_{cc}=\Op^2/\Os^2$.
This is again in congruence with the predictions in~\cite{kozlov2006coherence}, where the interpretation and physical mechanism of such a phenomenon are discussed at length.  By contrast, for a large detuning, i.e. $\Delta\gg\gamma,\Op,\Os$, we find that, as expected, excited populations are small in the steady state, with $\rho_{aa,cc}\approx\Omega_{\rm{p,s}}^2/(4\Delta^2)$.  
Finally, in the steady state, we find that the weak probe limit, in which $\rho_{ab}\approx\Op(2\Delta+\i\gamma)/(4\Delta^2+\gamma^2+2\Os^2)$, performs quite strongly as an approximation for the coherence term $\rho_{ab}$ (see~\figref{weakprobe}).
Other such limiting cases to previous approximate results are reproducible from our more general results, as the reader may verify.  We find no steady-state population inversion under any set of parameters. 

\subsection{Physical Implications}

\begin{figure}
    \centering
    \includegraphics[width=\columnwidth]{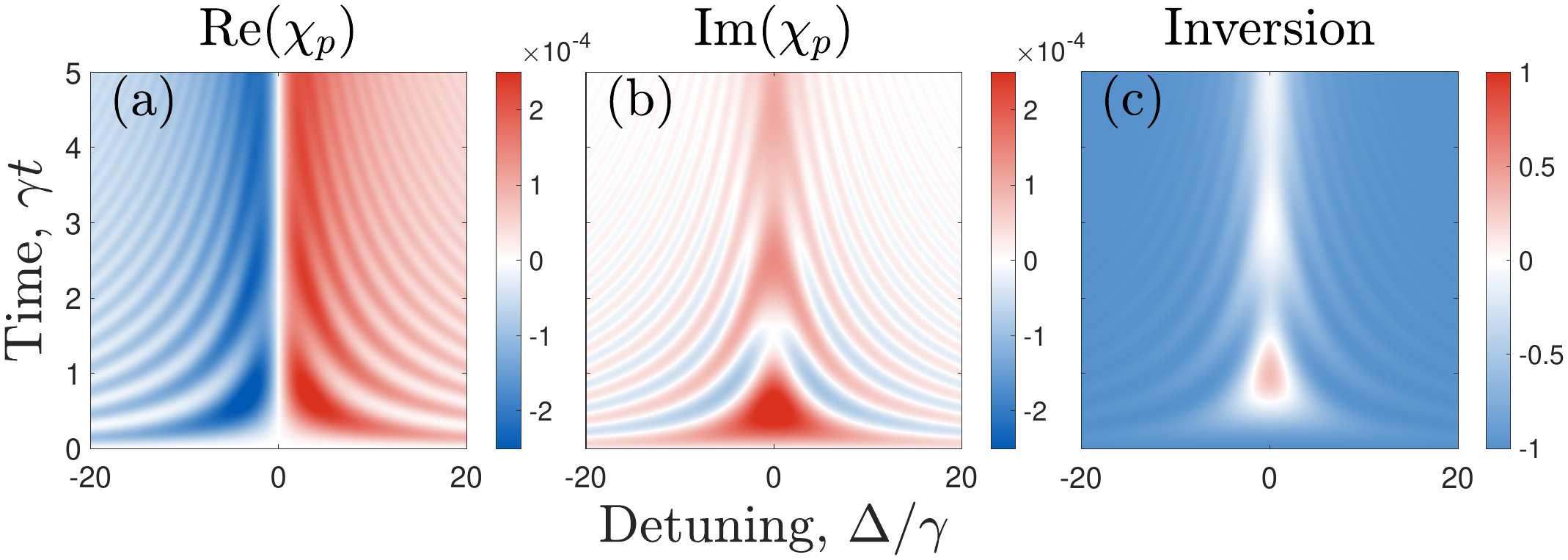}
    
    \caption{Transient (a) real and (b) imaginary parts of probe susceptibility representing the dispersion and absorption spectra, along with (c) transient population inversion, defined as $\rho_{aa}-\rho_{bb}$, all for the V-system. $\Im(\cp)>0$ corresponds to net absorption, and $\Im(\cp)<0$ corresponds to gain.  In (b) and (c) we observe instances of gain without inversion and inversion without gain.  Here, parameters have been chosen to mirror the Rubidium 87 system assembled in \cite{zibrov1995experiment} and are given in terms of the scaling term $\mu$, which is set to 1 kHz: $\Os=10^3\mu$, $\Op=3\Os$, $\gamma=10^3\mu$, $\omega_{ab}=10^9\mu$.}
    \label{VeeSusc}
\end{figure}

Generalizing our single system into a collection of 3-level systems, we now investigate optical effects of this collection on the driving field.
The frequency-dependent complex linear susceptibility of the medium in response to the probe field is calculated by
\begin{equation}\label{VSusc}
	\cp=\frac{2\mu}{\Op}\rho_{ab}(t),
\end{equation}
where $\mu$ is a scaling factor dependent on the physical implementation of the 3-level medium.  While qutrits may be realized by a variety of physical systems~\cite{blok2021qutrit} and we stress that our solution applies generally, 
for the purposes of a concrete example we consider the atomic gas, a common context in which EIT experiments are performed~\cite{zibrov1995experiment,kitching1999experiment}.  For an atomic gas, the factor is $\mu=\varrho|\mu_{ab}|^2/\hbar\epsilon_0$, where $\varrho$ is the atomic density.  In all plots, we give parameters in terms of $\mu$, but scaled according to experimental values in~\cite{zibrov1995experiment}, where a system of Rubidium 87 is assembled \cite{Steck2003Rubidium8D}.  Assuming a atomic density of $\sim 10^9 \textrm{cm}^{-3}$ such that $\mu\sim 1\rm{kHz}$, we ensure $\Op,\Os$, and $\gamma$ are $\mathscr{O}(\rm{MHz})$ and $\omega_{ab}$ is $\mathscr{O}(\rm{THz})$. Higher atomic densities such as those employed in~\cite{hau1999light} move the poles in the group velocity  to larger detunings (see below) but do not qualitatively change the results.  
Note that in deriving~\eqref{VSusc}, we have chosen the convention that $\Im(\cp)>0$ corresponds to net absorption of the probe field, while $\Im(\cp)<0$ corresponds to net gain~\cite{fleischhauer1992index}.
In \figref{VeeSusc}(a)-(b), we have plotted the transient real and imaginary parts of $\cp$, which represent the dispersion and absorption spectrum of the medium.  As we compare (b) with the transient population inversion along the $\ket a\leftrightarrow\ket b$ transition plotted in (c), we note instances of both gain without inversion and inversion without gain under these parameters.

Turning our attention toward the real part of the susceptibility, we now analytically calculate index of refraction and the group velocity of a light pulse propagating through the medium via the dispersion relation.  For a small susceptibility (i.e.~$|\cp|\ll1$), we approximate the index of refraction as $n\approx\sqrt{1+\Re(\cp)}$, valid in the 3-level medium near resonance~\cite{scullyzubairy}.  In this regime, we find an index of refraction very close to unity due to the low atomic density.  However, with a larger density, the index can be extremely large, as $n\propto\sqrt\varrho$ for $|\chi_p|\gg1$.  

For example, we once again consider the Rubidium 87 atomic gas assembled in \cite{zibrov1995experiment}, which employed the $5^2S_{1/2}\to 5^2P_{1/2}$ transition.  Along this transition, $\mu_{ab}=2.538\times10^{-29}C\cdot m$ and $\gamma=36.129$~MHz~\cite{Steck2003Rubidium8D}.  In a gas of such atoms with an atomic density of $\varrho=10^{14} \textrm{cm}^{-3}$ with driving parameters as in \figref{VeeSusc}, we predict the real part of the index of refraction to reach a maximum of 3.934.

\begin{figure}[b]
    \centering
    \includegraphics[width=\columnwidth]{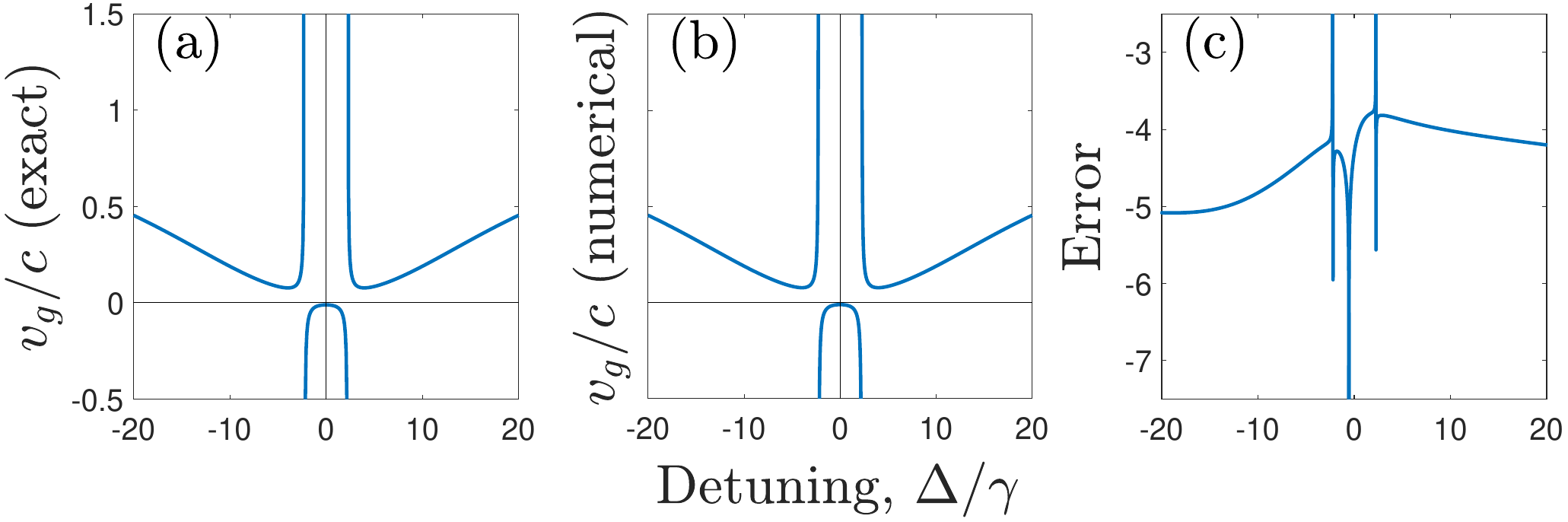}
    \caption{(a) Calculated exact steady-state group velocity for the probe field in units of $c$, with (b) the same calculated numerically and (c) the error of that numerical solution, calculated as $\log|(v_{g,exact}-v_{g,num.})/v_{g,exact}|$.  We observe superluminal group velocity near the poles at a detuning of $\simeq \pm 2\gamma$; the poles move by parameter choice (see~\eqref{VC}).  Our numerical results are obtained by performing the matrix exponential in \eqref{soln} numerically, then using MATLAB's gradient function with a resolution of $2\times10^4$ to compute the derivative.  Parameters are identical to those in \figref{VeeSusc}.}
    \label{VeeVC}
\end{figure}

Returning to our analytical results, we next calculate the steady-state group velocity (in units of the speed of light $c$) via the usual relation ${v_g}/{c}=\left(n+\omega_{ab}{\p n}/{\p \nu_{p}}\right)^{-1}$, obtaining \begin{equation}
    \frac{v_g}{c}=\frac{(4\Delta^2+\Gamma^2)(4\Delta^2+\Gamma^2+4\mu\Delta)}{(4\Delta^2+\Gamma^2)^2+8\mu\Delta\Gamma^2+2\mu(\Delta+\omega_{ab})(4\Delta^2-\Gamma^2)},
\label{VC}\end{equation} where $\Gamma=\sqrt{\gamma^2+2(\Op^2+\Os^2)}$. We note that the group velocity diverges with poles identifiable by the denominator in~\eqref{VC}.  In \figref{VeeVC}(a), we plot the calculated steady-state group velocity.  Near resonance, the figure displays both subluminal and superluminal group velocities, the latter found near the poles.  We even predict negative group velocities as the detuning nears exact resonance.  Also plotted in \figref{VeeVC}(b) is the same quantity calculated by a numerical simulation.  In (c), we compare the precision our analytical solution compared to that of a numerical solution, from which we see that the numerical solution is especially weak near the poles.
We also find that our expression for group velocity in~\eqref{VC} has potential zeroes at detunings given by \begin{equation}
    \Delta =-\frac12\Big(\mu \pm \sqrt{\mu^2-\Gamma^2} \Big).
\end{equation}
Thus, we predict the vanishing of the group velocity in the steady state at these detuning values.  Note, however, that these values are only real for ${\mu>\Gamma}$, i.e.~high atomic densities, and are thus not visible in~\figref{VeeVC}(a), where $\mu\ll\Gamma$.

However, there is a difficulty in measuring these steady-state effects experimentally.  The 3-level system does not support continuous EIT, which we observe by inspecting the imaginary part of the susceptibility. Under any parameters, we find a steady-state absorption spectrum with a Lorentzian lineshape in $\Delta$ given by
\begin{equation}
	\Im(\chi_{p,SS})=\frac{2\mu\gamma}{4\Delta^2+\gamma^2+2(\Op^2+\Os^2)}.
\end{equation}
Thus, the near-resonance refractive effects determined above will be met with large near-resonance absorption.  Experimental realizations of anomalous propagation in a V-system therefore have two options to avoid this difficulty.  First, one could measure group velocity before the steady state is reached, making use of the transient transparency windows found when $\gamma$ is on or below the order of the driving parameters $\Op,\Os$ (see \figref{VeeSusc}(b)).  We leave the calculation of transient group velocity for later consideration, though it is likely to yield similar behavior to the steady-state velocity due to the relative homogeneity of $\Re(\cp)$.  As an alternative, one could make use of a more complex multilevel system which contains a 3-level subsystem, such as the atoms used in~\cite{zibrov1995experiment,kitching1999experiment}.  These experiments utilized incoherent pumping into the auxiliary levels, enabling continuous EIT behavior, though they did not attempt to simultaneously observe the dispersive effects described here.  Slow-light experiments generally make use of systems in the $\Lambda$ configuration~\cite{hau1999slowlight,kasapi1995SLexperiment,phillips2001slowlight} to avoid these difficulties.

\section{\texorpdfstring{$\Lambda$-System}{Lambda-System}}
\label{sec:L}

\subsection{Analytical Results}

For the $\Lambda$-system initially in the $\ket b$ state, we obtain the following results for the density matrix populations:
\begin{align}
    \rho_{aa}(t)&=\frac{\Os^2}{\Omega^2}+\sum_{k=1}^4\left(\frac{A_k\exp(\lambda_kt)}{\prod_{l\neq k}^4(\lambda_k-\lambda_l)}\right)\\
    \rho_{bb}(t)&=\frac12\sum_{k=1}^4\left(\frac{B_k\exp(\lambda_kt)}{\prod_{l\neq k}^4(\lambda_k-\lambda_l)}\right)\\
    \rho_{cc}(t)&=1-\rho_{aa}(t)-\rho_{bb}(t)
\end{align}
where the $\lambda_k$ are the roots of the fourth degree polynomial
$\lambda(2\gamma+\lambda)(2\Delta^2+2(\gamma+\lambda)^2)+\Omega^2(\gamma+\lambda)(\gamma+2\lambda)$,
and we define
\begin{align}
    \Omega&=\sqrt{\Op^2+\Os^2}\\
    A_k&=\frac{\gamma\lambda_k\big(2\Delta^2+\Omega^2+2(\gamma+\lambda_k)^2\big)+\Op^2\lambda_k^2+\Os^2\gamma^2}{2\lambda_k}\\
    B_k&=2\Delta^2\lambda_k+\Omega^2(\gamma+\lambda_k)+2\lambda_k(\gamma+\lambda_k)^2.
\end{align}

For the coherence elements, we find
\begin{align}
    \rho_{ab}(t)&=\frac{\Op}2\sum_{k=1}^4\left(\frac{\left(\Delta(\gamma-\lambda_k)+\i(\lambda_k^2-\gamma^2)\right)\exp(\lambda_kt)}{2\prod_{l\neq k}^4(\lambda_k-\lambda_l)}\right)\\
    \rho_{bc}(t)&=\frac{\Os}{\Op}\rho_{ab}^*(t)\\
    \rho_{ac}(t)&=-\frac{\Op\Os}{\Omega^2}+\sum_{k=1}^4\left(\frac{C_k\exp(\lambda_kt)}{\prod_{l\neq k}^4(\lambda_k-\lambda_l)}\right),
\end{align}
with $C_k=\Op\Os(\lambda_k^2-\gamma^2)/2\lambda_k$.  Again, the density matrix elements consist of both constant terms and time-dependent terms.  In the steady-state, these time-dependent terms vanish, leaving only the constant terms.

\subsection{Physical Implications}

\begin{figure}[t]
    \centering
    \includegraphics[width=\columnwidth]{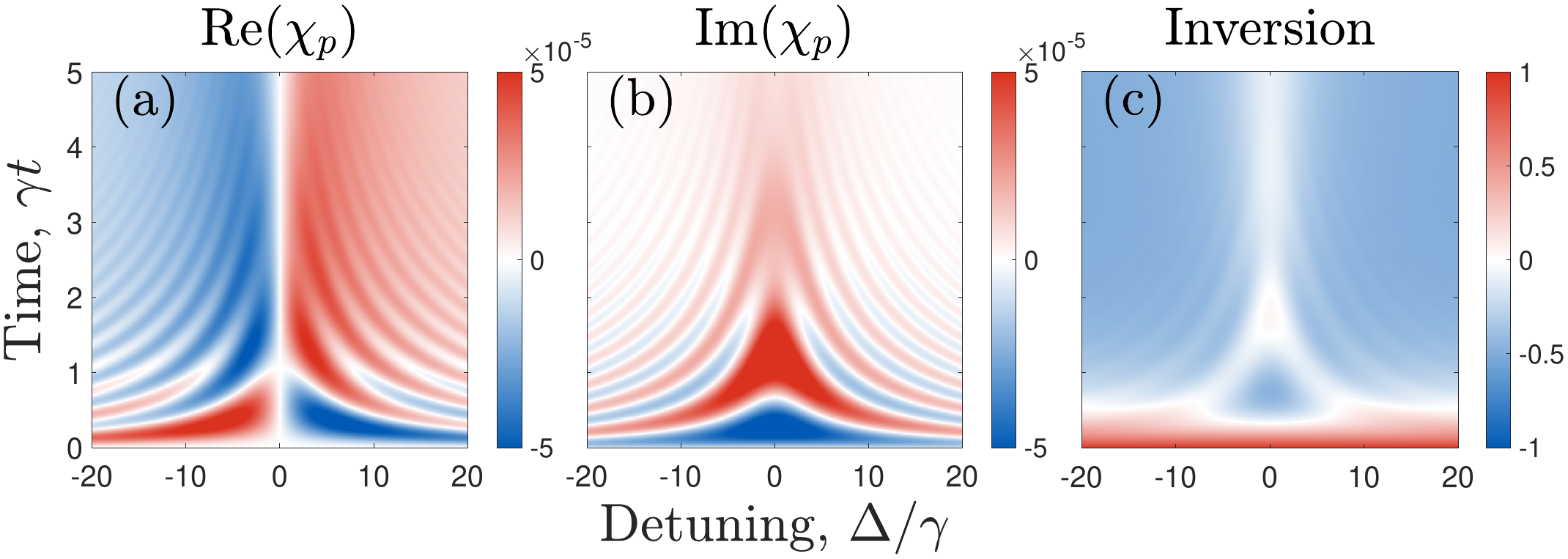}
    \caption{Transient (a) real and (b) imaginary parts of the linear susceptibility in response to the probe field, along with (c) population inversion along the probe transition, defined as $\rho_{bb}-\rho_{aa}$, all in the $\Lambda$-system.  Comparisons between (b) and (c) display intense gain where inversion is present, and again display regimes of gain without inversion and inversion without gain.  Parameters are identical to those in \figref{VeeSusc}}
    \label{LambdaSusc}
\end{figure}

We first note that under any parameters, the population of $\ket b$ eventually decays to zero, and the only nonzero terms in the steady state are\begin{align}\label{Lss}
    \rho_{aa}&=\frac{\Os^2}{\Omega^2} & \rho_{cc}&=\frac{\Op^2}{\Omega^2} & \rho_{ac}&=-\frac{\Op\Os}{\Omega^2}.
\end{align}
Somewhat counter-intuitively, the ratio of populations in the $\ket a$ and $\ket c$ states is $\rho_{aa}/\rho_{cc}=\Os^2/\Op^2$, i.e. stronger driving on the $\ket c$ transition increases the final population of $\ket a$.  In contrast to the mixed state we reach in the V-system,~\eqref{Lss} represents a pure state, and we conclude that the final quantum state of the system is a multiple of $(\Os\ket a -\Op\ket c)/\sqrt{\Op^2+\Os^2}$.  This state is identical to the non-coupled ``dark state'' superposition identified in~\cite{arimondo1996CPT}.  Absorption and consequent excitation of atoms from this state is not possible, and the population is effectively trapped in this coherent state, even under continuous driving.

A direct consequence of this is that there is no coherence supported along the probe transition in the steady state. Although the susceptibility will vanish and the index of refraction will approach unity in the steady state, we still observe interesting transient effects.  In the $\Lambda$ system, the susceptibility in response to the probe field is given by \begin{equation}
    \cp=\frac{2\mu}{\Op}\rho_{ba}(t)
\end{equation} (note the difference from the susceptibility in the V-system in~\eqref{VSusc}).  In the transient regime, we again find instances of anomalous dispersion and gain without inversion (see \figref{LambdaSusc}).  In particular, the real part of the susceptibility in \ref{LambdaSusc}(a) is highly irregular, with the derivative at $\Delta=0$ reversing signs near $\gamma t= 1$.  While this should make for intriguing transient group velocity behaviour, we leave this calculation for future consideration.

\section{Conclusions}\label{sec:conc}

We found the exact analytical form of the dynamics of the driven qutrit in the V and $\Lambda$ configurations using the Lindlbad master equation.  This exact solution gave us new insight into transient regimes of LWI and EIT in these systems through a calculation of the linear susceptibility; identified parameters for steady-state superluminal, vanishing, and negative group velocities in the V-system; and found an inevitable decay into the population trapped state in the $\Lambda$-system. 

These exact solutions allow for precise preparation of 3-level subsystems of the complex multilevel quantum systems required for supporting continuous lasing without inversion (LWI) and electromagnetically induced transparency (EIT), as well as for observing the dispersive effects we have identified here.  Beyond these demonstrative optical effects, precise preparations of multilevel atoms is a crucial step in the quantum computing process.  Using our solution, under Lindblad conditions, a V-type atom can be prepared in nearly any mixed state superposition by precise selection of the system parameters, and a $\Lambda$-type atom can be prepared in any pure state superposition of its lower states.  

In future work, the methods of our solution may be extended to other 3- or higher-level systems.  Moreover, our exact solution can be leveraged to determine qutrit decoherence under various quantum control strategies for quantum computing and other quantum technology applications.  Trapped ion quantum simulator~\cite{altman2021quantum} and quantum computing~\cite{alexeev2021quantum} architectures in particular, as well as quantum testbeds at Berkeley~\cite{blok2021qutrit,berkeley2021qutrit} and Sandia National Labs~\cite{sandia2021qutrit} have taken advantage of qutrit quantum computing scenarios~\cite{low2020qutrit} and may benefit from the results of this work.


\section*{Acknowledgements}
    We acknowledge useful conversations with Charles Durfee, Daniel Jaschke, Matthew Jones, Nathan Smith, and Gavriil Shchedrin.  This work was performed in part with support by the NSF
    under grants OAC-1740130, CCF-1839232, PHY-1806372; and in conjunction with
    the QSUM program, which is supported by the Engineering and Physical Sciences
    Research Council grant EP/P01058X/1.

\bibliography{bibliography.bib}

\end{document}